\documentstyle[prl,aps,epsf]{revtex}

\begin{document}

\title{Bipolaronic charge excitations in t-J two-leg ladders}
\author{D. A. Ivanov and Patrick A. Lee}
\address{Department of Physics,
Massachusetts Institute of Technology,
Cambridge, Massachusetts 02139, USA}
\date{August 18, 1997}
\maketitle

\begin{abstract}
We present a low-energy effective model for the charge
degrees of freedom in two-leg t-J ladders. Starting 
from $SU(2)$ mean-field theory, we exclude the spin degrees
of freedom which have an energy gap. At low temperatures,
the mean-field solution is the staggered-flux phase.
For gapless charge excitations the effective theory
is the Luther-Emery liquid. Our analysis is applicable at low
doping and in the ``physical'' range of parameters $t/J\sim 3$
where there is only one massless mode in the charge sector
and no massless modes in the spin sector. Within our model
we make predictions about correlation exponents and the
superconductivity order parameter, and discuss the 
comparison with the existing numerical results.
\end{abstract}


\section{Introduction}

Theoretical studies of t-J ladders have proven to be
valuable for understanding Hi-$T_c$ cuprate compounds. While
containing certain features of the two-dimensional t-J
model, the ladders are quasi-one-dimensional, which
greatly simplifies the treatment of the problem. Recently,
cuprate compounds with ladder structures have been produced
experimentally, and data on their electronic properties have
been obtained \cite{exp}. The ladder compounds may be used for
verification of models and mechanisms of superconductivity
proposed for layered cuprates. The experimental works are
now complemented by numerical results on t-J 
and Hubbard models \cite{NWS,SRZ}. Together with experimental 
results, they provide a good testing ground for any analytical
treatment.

The challenge of theoretically solving the t-J model on ladders
arises from its strongly correlated nature. In the real ladder
compounds the coupling is nearly isotropic, i.e. the interchain
coupling parameters ($t$ and $J$) are close to the intrachain
coupling. Thus, the problem does not have a small parameter and
cannot be treated by a standard perturbation theory. Several
works exist based on starting from uncoupled chains and then
including interchain hopping and spin exchange as perturbations 
\cite{Kh,N}. While weak-coupling approach
is the most consistent and controlled of the existing
analytic methods, it is not completely reliable as the
interchain coupling increases and
approaches the single-chain bandwidth. We shall
further comment on possible corrections to this treatment.

In the present paper we employ the $SU(2)$ slave-boson mean-field
approach \cite{WL,LNNW}. Although not a controlled approximation, 
we believe that it can
correctly capture the low-energy physics of the systems with a
spin gap. In the paper we specialize to two-leg ladders, but
our treatment may be further extended to any even-leg ladders which are
known to exhibit spin gap. 
The spin gap ensures that most of the fluctuations around the mean-field
state are massive. The only massless fluctuation is the 1+1-dimensional
abelian gauge field which can be explicitly included as a pair-binding
potential.

The general idea of slave-boson method is to represent the vacant
sites (holes) by an auxiliary bosonic field, which allows us to rewrite
the non-linear no-double-occupancy condition as a linear constraint
in terms of fermions (representing spin degrees of freedom) and
bosons (representing charge degrees of freedom) \cite{WL,LNNW,LN}.
Introducing auxiliary bosons expands the Hilbert space of states, and
the system aquires an additional gauge symmetry. The mean-field
ansatz breaks this symmetry, which is restored for physical correlation
functions after averaging over all gauge-equivalent 
configurations.

We use the $SU(2)$ version of the slave-boson construction developed
earlier for the two-dimensional problem \cite{WL,LNNW}. In this method, the
auxiliary boson has two components (we call the corresponding degree
of freedom isospin; it is distinct from the actual spin) which describe
holes in the two different ways: either as sites with no fermions or
as sites doubly occupied by fermions. Thus extended, the system
has a $SU(2)$ gauge symmetry (rotating isospin).
We choose to use the $SU(2)$ formalism instead of $U(1)$ version
of the slave-boson method developed previously for two-dimensional
t-J model \cite{LN,UL}.
We note that both analytic and numerical works \cite{NWS,N,TTR} point to
a bipolaron picture where the holes are bound in pairs. 
As we shall see, this
picture naturally emerges out of the $SU(2)$ formulation as the
confinement between two species of bosons, while the $U(1)$ formalism
fails to give the correct low-energy physics. We believe that
the $SU(2)$ mean-field theory has an advantage at low doping where
it generalizes the $SU(2)$ symmetry of t-J model at half-filling \cite{WL}.
A discussion of
relation between $U(1)$ and $SU(2)$ approaches may be found in \cite{LNNW}.

We find that the low-temperature mean-field phase is the 
staggered-flux phase (similar to that found for the two-dimensional problem
in \cite{WL,LNNW}). The fluctuations about this mean-field state are
described by gauge fields which have a gap, except for the in-phase
fluctuations which form a 1+1-dimensional abelian gauge field. As a result,
we find that the low-energy effective theory consists of two
degenerate bands of holons with a short-range interaction and coupled to
a $U(1)$ gauge-field. The dynamics of this gauge field arises from
its interactions with spinons. Spinons have a gap and, therefore,
give a nonsingular  dynamics to the gauge field with the energy scale
$J$. The two bands of holons have opposite charges with respect to
this gauge field and, therefore, form confined pairs.
This leads us to the conclusion that the resulting theory
for the hole excitations is the Luther-Emery liquid
of hole pairs. This agrees with the
bipolaronic picture of charge excitations proposed earlier both
analytically and numerically \cite{NWS,N,TTR}. It is known that the 
Luther-Emery liquid has two competing orders: superconducting singlet
pairing (SS) and charge density wave (CDW) \cite{Kh,N,EL}. We point out the
necessity to distinguish between the hole density (two-particle
operator) and the pair density (four-particle operator). While
the product of the correlation exponents for the SS and CDW order
parameters is equal to one when CDW is understood as hole-density
correlations, this relation does not necessarily hold 
for four-particle pair-density correlations. 
The relation between the single-hole and pair CDW exponents depends
on the degree of the overlap of the bipolaronic pairs. At low density
of holes, when the pairs do not overlap, these two exponents coincide.
On the other hand, in the limit of highly overlapping pairs we find
that the effective exponents may differ by 2.
This possibly explains the unexpected numerical
results for the correlation exponents obtained by Noack et al. \cite{NWS}.

Further, we discuss the possible implication of our model for
the superconducting transition via pair condensation (of course,
interladder correlations would be necessary). We describe the
superconducting order by the nearest-neighbor order parameter
$\Delta_{ij}$. We extend our discussion for a more general case
of a weakly doped antiferromagnet on a bi-partite lattice 
with a spin gap. This class of systems includes all even-leg ladders
as a particular case.
We assume that at low temperature such a system
is in the staggered flux phase, which results in two degenerate
interacting holonic bands. Under these assupmtions we find that
the order parameter obeys the modified d-wave relation:
\begin{equation}
\sum_j t_{ij}\Delta_{ij}=0, 
\label{1.1}
\end{equation}
where the sum is performed over all nearest neighbors of a site $i$.
This relation holds in the limit of zero doping, with
corrections involving the hole concentration.
This relation was first derived by S.~C.~Zhang as an exact
result for the Hubbard model \cite{Z}. Our derivation should be
understood as a verification that the $SU(2)$ mean-field approximation
preserves this exact property. When specialized
to the case of the two-leg ladder with isotropic coupling,
the above equation becomes
$\Delta_\perp=-2\Delta_\parallel$. This agrees very well with 
the earlier numerical results \cite{NWS,SRZ}.

The rest of the paper is organized in three sections. In the first
section, we review the $SU(2)$ slave-boson method and present the
results of the mean-field theory computations. In the second section
we discuss the effective theory for the holons and the correlation
exponents for SS and CDW pairing. Finally, the third part is
devoted to the discussion of the modified d-wave relation for the
superconducting order parameter.

\section{$SU(2)$ mean-field theory of the ladder}

\begin{figure}
\centerline{\epsffile{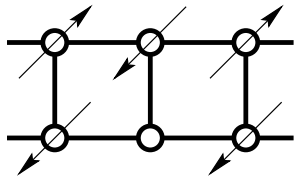}}
\caption{Two-leg t-J ladder.}
\label{fig:1}
\end{figure}

In this section we present the $SU(2)$ mean-field theory for
the t-J Hamiltonian
\begin{equation}
H=\sum_{\{ ij\} } J (\vec S_i \vec S_j
-{1\over4}n_i n_j) -t{\cal P} (c_{\alpha i}^\dagger c_{\alpha j}
+{\rm h.c.}){\cal P}
\label{2.1}
\end{equation}
on the ladder (Fig.~\ref{fig:1}). 


The sum is performed over nearest-neighbor site
pairs $\{ i,j \}$, $c_{\alpha i}^\dagger$ and $c_{\alpha i}$ are the
electron creation and annihilation operators on site $i$ ($\alpha$
is the spin index), $\vec S_i$ is the electron spin,
$\vec S_i={1\over2} c_{\alpha i}^\dagger \vec \sigma_{\alpha\beta}
c_{\beta i}$, $n_i$ is the occupation nuber of the site $i$
($n_i=c_{\alpha i}^\dagger c_{\alpha i}$), $\cal P$ is the
projector onto the no-double-occupancy states (with $n_i\le 1$ for any
$i$). $t$ and $J$ are the parameters of the Hamiltonian. In the real
ladder compounds $t$ and $J$ are estimated to be about 4000K and
1300K respectively \cite{exp}. In what follows we assume that the interchain
and the intrachain couplings are equal ($t_\perp = t_\parallel=t$,
$J_\perp = J_\parallel = J$) and as a realistic approximation we take
$t/J=3$.

Following the usual procedure of the $SU(2)$ slave-boson method 
\cite{WL,LNNW},
we introduce two fermionic and one bosonic isospin doublets on each
site:
\begin{equation}
\psi_{1i}=\pmatrix{f_{1i}\cr f_{2i}^\dagger}
\qquad \psi_{2i}=\pmatrix{f_{2i}\cr -f_{1i}^\dagger}
\qquad h_i=\pmatrix{b_{1i} \cr b_{2i}}
\end{equation}
with the electronic operators written in terms of bosons
$h_i$ and fermions $\psi_{\alpha i}$ as
\begin{equation}
c_{\alpha i}={1\over\sqrt2} h_i^\dagger\psi_{\alpha i}
\end{equation}
The resulting Hilbert space is larger than that of the original t-J
system. To select the subspace of physical states (which is invariant
under the Hamiltonian of the original system) we impose a linear
constraint (replacing the non-linear no-double-occupancy constraint):
\begin{equation}
({1\over2}\psi_{\alpha i}^\dagger \vec \tau \psi_{\alpha i}
+h_i^\dagger \vec \tau h_i )|phys\rangle =0
\label{2.4}
\end{equation}
($\vec \tau$ are identical to the Pauli matrices $\vec \sigma$, but
they act in the isospin space, and we denote them by a different
letter to distinguish from the Pauli matrices $\vec \sigma$ acting
on true spin). On a given site, this constraint allows only three
states: $f_1^\dagger |0\rangle$, $f_2^\dagger|0\rangle$, and
${1\over\sqrt2}(b_1^\dagger+b_2^\dagger f_1^\dagger f_2^\dagger)|0\rangle$,
which correspond to spin up, spin down electrons and a vacancy respectively.
Thus in the $SU(2)$ formulation, a vacancy may be represented by both
two-spinon and no-spinon states corresponding to different isospins
of the holon $h_i$.

Thus formulated, the extended system is invariant under an $SU(2)$ gauge
symmetry:
\begin{equation}
\psi_{\alpha i} \mapsto g_i \psi_{\alpha i},
\qquad h_i \mapsto g_i h_i. 
\end{equation}
This gauge symmetry acts on the isospin of fermions and bosons, and
mixes creation operators $f_{\alpha i}^\dagger$ with the annihilation
operators of opposite spin $f_{-\alpha i}$.

Introducing the nearest-neighbor mean-field parameters 
\begin{equation}
U_{ij}=\langle\psi_{\alpha i} \psi_{\alpha j}^\dagger \rangle -
{t\over J} (\langle h_i h_j^\dagger \rangle +{\bf I} \langle
h_j h_i^\dagger \rangle^T {\bf I} )
\label{2.6}
\end{equation}
with
\begin{equation}
{\bf I}=\pmatrix {0 & 1 \cr -1 & 0}
\end{equation}
and the Lagrange multipliers $a_i^\mu$ ($\mu=1,2,3$)
for enforcing the linear constraint (\ref{2.4}), the
mean-field Hamiltonian becomes
\begin{equation}
H=\sum_{\{ ij\} }\left[
{J\over 4} {\rm Tr}(U_{ij}U_{ij}^\dagger) + {J\over2}\psi_{\alpha i}^\dagger
U_{ij}\psi_{\alpha j} +{t\over2} (h_i^\dagger U_{ij} h_j + {\rm h.c.})
\right] +\sum_i a_i^\mu ({1\over2}\psi_{\alpha i}^\dagger
\tau_\mu \psi_{\alpha i} + h_i^\dagger \tau_\mu h_i).
\label{2.8}
\end{equation}

In the present paper we use combinatoric coefficients in (\ref{2.8}) different
from those used in \cite{WL,LNNW,UL}. 
Our choice of coefficients in (\ref{2.8}) gives the
correct combinatoric factors for tadpole diagrams,
and they differ from those obtained from the Hubbard-Stratonovich
transformation. This ambiguity in the numeric factors of order one
is of no practical significance, since it falls 
within the uncertainty of the mean-field
approximation.

The mean-field Hamiltonian (\ref{2.8}) is accurate only up to four-boson
terms which we neglect in the usual way \cite{LN}.
The four-boson terms give rise to only short-range interactions
between holons and can be omitted at this level of approximation.

In \cite{WL} it was argued that the non-zero values of $a_i^\mu$ correspond
to Bose condensation. We do not expect Bose condensation in a
quasi-one-dimensional system and set mean-field value $a_i^\mu=0$.
This implies that we in fact release the no-double-occupancy
constraint (it is satisfied only on average). 
Afterwards, the constraint may be imposed by including the
fluctuations of the field $a_i^\mu$.

Introducing the chemical potential $\mu$, we arrive at the mean-field
Hamiltonian
\begin{equation}
H=\sum_{\{ ij\} } \left[ {J\over4} {\rm Tr} (U_{ij} U_{ij}^\dagger)
+{J\over2} \psi_{\alpha i}^\dagger U_{ij} \psi_{\alpha j}
+{t\over2} (h_i^\dagger U_{ij} h_j + {\rm h.c.})\right]
-\mu\sum_i(h_i^\dagger h_i -\delta),
\label{2.9}
\end{equation}
where $\delta$ is the concentration of holes (doping). The matrices $U_{ij}$
have the form
\begin{equation}
U_{ij}=\pmatrix{\chi_{ij} & \Delta_{ij} \cr \Delta_{ij}^* &
-\chi_{ij}^* } = U_{ji}^\dagger = i a_{ij} G_{ij},
\end{equation}
where $a_{ij}$ are positive real numbers (amplitudes),
$G_{ij}\in SU(2)$ are $2\times 2$ matrices.

The mean-field saddle point at temperature $T$ is found as the
extremum of the free energy
\begin{equation}
F[U_{ij}]=-T\log {\rm Tr}_{h,\psi} \exp(-H/T).
\label{2.11}
\end{equation}
$F[U_{ij}]$ is invariant under $SU(2)$ gauge transformations
\begin{equation}
U_{ij}\mapsto W_i U_{ij} W_j^\dagger
\end{equation}
for any set of $SU(2)$ elements $W_i$.

The mean-field solution breaks this gauge symmetry. Only gauge
invariant quantitites correspond to physical observables. Any
non-gauge-invariant expression will vanish after averaging over
all gauge-equivalent configurations of $U_{ij}$. 

\begin{figure}
\centerline{\epsffile{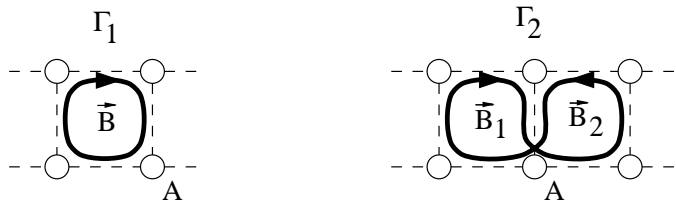}}
\caption{The two closed contours used in constructing 
mean-field order parameters.}
\label{fig:2}
\end{figure}

Now we turn to describing possible phases. Phases should be parametrized
by gauge-invariant functions of $U_{ij}$. We assume that the translational
symmetry is unbroken, i.e. a translation of the mean-field solution
$\{ U_{ij} \}$ along the ladder transforms it to a gauge equivalent
configuration. We also assume that the symmetry of reflection about
the ladder axis (interchanging the two legs) is also preserved in the
mean-field solution. Under these assumptions all possible phases may be
parametrized by four real parameters: the two amplitudes
$a_\parallel$ and $a_\perp$ (intrachain and interchain respectively),
and two $SU(2)$ order parameters:
\begin{equation}
b={1\over2} {\rm Tr} \prod_{\Gamma_1} G_{ij},
\end{equation}
\begin{equation}
c={1\over2} {\rm Tr} \prod_{\Gamma_2} G_{ij}
\end{equation}
with the products taken along the contours $\Gamma_1$ and $\Gamma_2$
shown in Fig.~\ref{fig:2} (the first product contains four matrices, the
second one --- eight matrices).


The meaning of the order parameter $b$ is analogous to the cosine
of the flux through plaquet in the $U(1)$ formulation \cite{LN}. To explain
this analogy we may introduce the $SU(2)$ flux $\vec B$ defined by
\begin{equation} 
\exp(i\vec B \cdot \vec \tau)=\prod_{\Gamma_1} G_{ij},
\end{equation}
where the product starts and ends at a site $A$ of the contour
$\Gamma_1$ (Fig.~\ref{fig:2}). Then $b=\cos |\vec B|$ (obviously, the direction
of the vector $\vec B$ depends on the choice of the starting point
$A$, but its magnitude $|\vec B|$ does not). 

\begin{figure}
\centerline{\epsffile{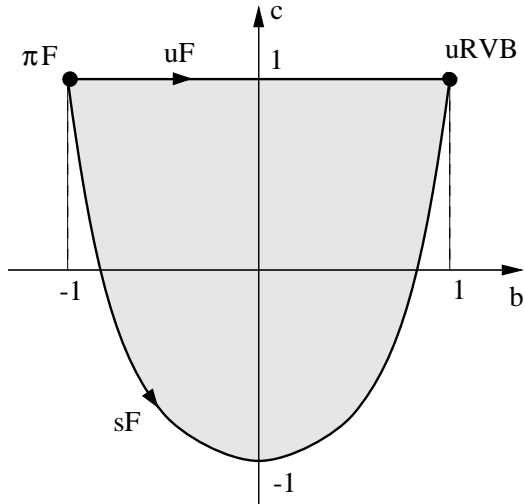}}
\caption{Space of mean-field phases. The two corners correspond to the
$\pi$-flux and uRVB phases, boundaries --- to the uniform-flux and
staggered-flux phases.}
\label{fig:3}
\end{figure}

The order parameter
$c$ measures the relative orientation of neighboring
$SU(2)$ fluxes. Namely, define the two fluxes $\vec B_1$ and 
$\vec B_2$ through neighboring plaquets with a common starting
point $A$ (Fig.~\ref{fig:2}). Then
\begin{equation}
c={1\over2} {\rm Tr} \left(\exp(i\vec B_1 \cdot \vec \tau) 
\exp(-i\vec B_2 \cdot \vec \tau) \right)
=\cos|\vec B_1|\cos|\vec B_2|+ {(\vec B_1 \cdot \vec B_2)\over
|\vec B_1| |\vec B_2|}
\sin|\vec B_1|\sin|\vec B_2|.
\end{equation}
Then we see that for any $G_{ij}\in SU(2)$
the order parameters $b$ and $c$ are restricted to
\begin{equation}
 -1\le b \le 1,
\end{equation}
\begin{equation}
2b^2-1 \le c \le 1.
\end{equation}
Thus, the $SU(2)$ order may be represented by a point in a two-dimensional
domain (Fig.~\ref{fig:3}) with the two corners representing the $\pi$-flux and
the uRVB phases, the boundaries corresponding to the uniform flux (uF)
and staggered flux (sF) phases analogous to their $U(1)$ counterparts 
\cite{UL},
but preserving translational and time-reversal symmetries.
In the uF phase the neighboring $SU(2)$ fluxes are parallel, in the
sF phase they are antiparallel, and inside the shaded region in 
Fig.~\ref{fig:3}
they form angles ranging between 0 and $\pi$.


By numerically minimizing the free energy (\ref{2.11}) (at $t/J=3$)
we find the following mean-field phase diagram in the $(\delta,T)$ coordinates:


\begin{figure}
\centerline{\epsffile{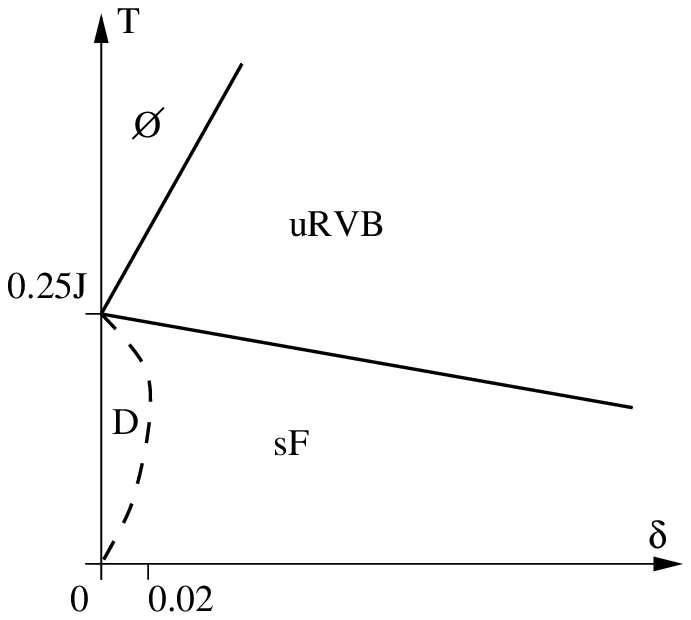}}
\caption{Mean-field phasee diagram at isotropic coupling and $t/J=3$.}
\label{fig:4}
\end{figure}
 
where
\begin{enumerate}
\item 
$\emptyset$ denotes the high-temperature free spin phase
($a_\parallel=a_\perp=0$);

\item 
uRVB is the phase with $a_\parallel \ne 0$, $a_\perp \ne 0$,
$G_{ij}={\bf 1}$ (so that $b=c=1$);

\item
D is the dimer phase: $a_\perp \ne 0$, $a_\parallel=0$;

\item
sF is the staggered-flux phase with $a_\perp \ne 0$,
$a_\parallel \ne 0$, $-1<b<1$, $c=2b^2-1$.
\end{enumerate}

Numerically we find that
the transition between D and sF phases is a very soft first-order
transition. In fact, the dimer phase D has flat spectra for
bosons and fermions and will be destroyed by the fluctuations
(correlations along the ladder will appear). We shall
disregard the dimer phase as an artifact of the mean-field approximation
and for the rest of the paper we restrict our discussion to the sF phase.

\begin{figure}
\centerline{\epsffile{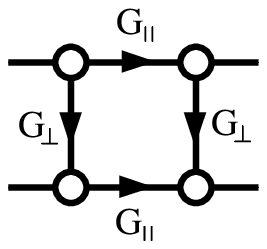}}
\caption{sF phase order parameters in the translationally invariant gauge.}
\label{fig:5}
\end{figure}

The sF phase may be described by different gauge-equivalent configurations
of $U_{ij}$. One of the {\it translationally invariant} configurations
(analogous to d-wave pairing phase in the $U(1)$ mean-field theory \cite{UL})
is (Fig.~\ref{fig:5}):
\begin{equation}
G_\perp=i\tau_1, \qquad G_\parallel=i(\cos{\varphi\over2}\tau_1
+\sin{\varphi\over2}\tau_2),
\end{equation}
or, equivalently,
\begin{equation}
U_\perp=\pmatrix{0 & a_\perp \cr a_\perp & 0}, \qquad
U_\parallel=\pmatrix{0 & a_\parallel e^{i{\varphi\over2}} \cr
a_\parallel e^{-i{\varphi\over2}} & 0}.
\end{equation}


\begin{figure}
\centerline{\epsffile{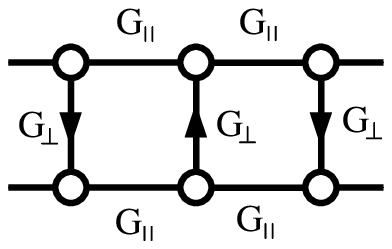}}
\caption{sF order parameters in the abelian gauge.}
\label{fig:6}
\end{figure}

Further, we shall use a different, the so called ``abelian'' parametrization
with
\begin{equation}
G_\parallel={\bf 1}, \qquad G_\perp=\cos{\varphi\over2}
+(-1)^m i\tau_3 \sin{\varphi\over2},
\label{2.21}
\end{equation}
(Fig.~\ref{fig:6}) which corresponds to
\begin{equation}
U_\parallel=i\pmatrix{a_\parallel & 0 \cr 0 & a_\parallel},
\qquad U_\perp=i\pmatrix{a_\perp e^{i (-1)^m {\varphi\over2}} & 0 \cr
0 &a_\perp  e^{-i(-1)^m {\varphi\over2}} },
\label{2.22}
\end{equation}
where $m$ is the number of the rung. This gauge fixing is not translationally
invariant, but instead it has the property that all $U_{ij}$ commute.
This choice of gauge resembles the staggered-flux $U(1)$ phase \cite{UL}.
In the U(1) formalism, the staggered-flux and d-wave pairing phases
are different, but their $SU(2)$ counterparts are gauge-equivalent \cite{WL}.


In the $SU(2)$ mean-field theory bosonic and fermionic spectra are
proportional to each other (with the scales $t$ and $J$ respectively).
In the sF phase we obtain the spectrum
\begin{equation}
{E_f\over J}={E_b\over t}=\pm{1\over2}\sqrt{a_\perp^2+
(2a_\parallel)^2 \cos^2 k + 2(2a_\parallel)a_\perp\cos{\varphi\over2}
\cos k}
\end{equation}
($k\in [0; 2\pi]$ is the wave vector) and each of these two bands is
doubly degenerate (four bands total). We remark that labeling the
states by wave vectors $k$ may depend on the gauge, but, because
gauge-dependent shift of $k$ involves equally bosons and fermions,
the gauge-invariant quantities remain unchanged.

The double degeneracy of states is a characteristic feature of the
staggered-flux phase. It is due to the fact that a certain non-abelian
subgroup of the full symmetry group of the extended Hamiltonian
remains unbroken by the mean-field ansatz. In the staggered-flux phase
the order parameters $U_{ij}$ are invariant under a U(1) subgroup of
global isospin rotations (in the abelian gauge (\ref{2.21}) -- (\ref{2.22}) 
these
are simply global rotations by $\tau_3$). Besides, there remains a particular
symmetry of $U_{ij}$, which is the combination of time reversal (transforming
$U_{ij}\mapsto U_{ij}^*$) and global isospin rotation (exchanging isospins
up and down in the abelian gauge). This symmetry operation does not
commute with the $U(1)$ rotation, but extends them to a nonabelian
group. The two degenerate bands form a two-dimensional representation
of this group, with time reversal mapping one band onto the other.

\begin{figure}
\centerline{\epsffile{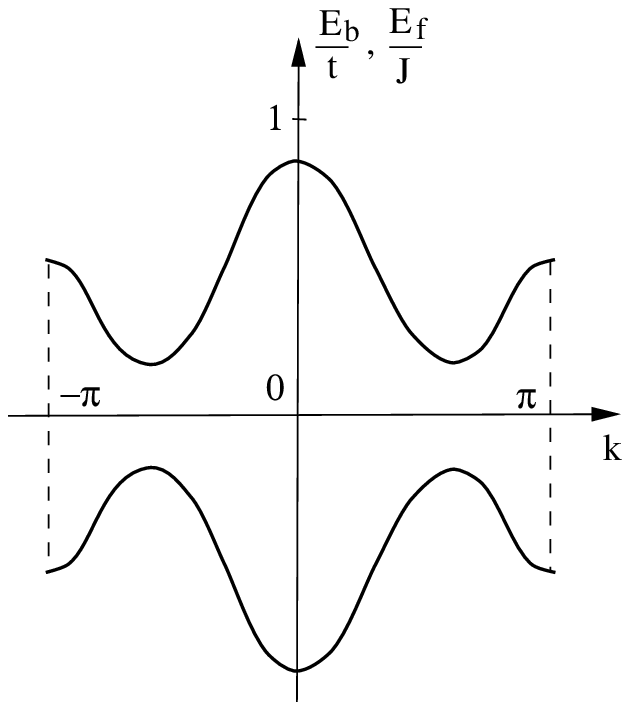}}
\caption{Typical fermionic-bosonic spectrum. 
Both bands are doubly degenerate.}
\label{fig:7}
\end{figure}

The typical numerical values for $a_\parallel$, $a_\perp$ and 
$\cos{\varphi\over2}$ are $a_\parallel=0.5$, $a_\perp=0.9$,
$\cos{\varphi\over2}=0.4$ (fonud by minimizing
free energy at $\delta=0.05$, $T=0.1J$, $t/J=3$).
This means that the upper bands of the spectrum are separated
from the lower bands by a gap of order $J$ for fermions
and of order $t$ for bosons (Fig.~\ref{fig:7}). The fermionic spectrum is
half-filled, i.e. the lower bands are completely filled whereas
the upper bands are empty. The fermionic excitations have a gap
of order of $J$. This agrees with the prediction of spin gap
in two-leg ladders and is crucial for the stability of the phase.


Finally, we need to include the fluctuations about the mean-field
phase. These fluctuations 
are described by the spatial $SU(2)$ gauge fields
$A_{ij}=U_{ij}^{-1}\delta U_{ij}$ defined on the links
and their temporal
counterparts $a_i=a_i^\mu \tau_\mu$ defined on the sites \cite{WL,LNNW,LN}. 
The temporal components of the gauge field $a_i$ coincide with
the constraint-fixing Lagrange multipliers in (\ref{2.8}) when
expressed in units of $J$. The effective action for the
fields $A_{ij}$ and $a_i$ is invariant under the (time-dependent) gauge
transformations
\begin{equation}
A_{ij} \mapsto W_i(t) A_{ij} W_j^\dagger(t), \qquad 
a_i \mapsto a_i + i \partial_t W_i(t)
\label{2.24}
\end{equation}
for arbitrary $SU(2)$ matrices $W_i(t)$ defined on sites.

In the sF phase the $SU(2)$
gauge symmetry is broken down to U(1) global symmetry. In the abelian
gauge, this residual symmetry is realized by rotations by $\tau_3$. As
shown in the Appendix B of \cite{LNNW}, the free energy will contain
terms proportional to ${\rm Tr}\prod U_{ij}$, where the products are taken
along closed loops on the lattice. When expanded in gauge-field
fluctuations, these terms give rise to mass for the gauge modes
proportional to $\tau_1$ and $\tau_2$. Thus, massless modes of the
gauge field must be proportional to $\tau_3$ \cite{LNNW}, and we may
treat $A_{ij}$ and $a_i$ as $U(1)$ gauge fields.

\begin{figure}
\centerline{\epsffile{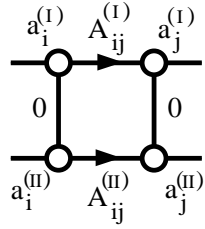}}
\caption{Gauge fields describing the fluctuations of the order parameter.
We choose the gauge with $A_\perp=0$.}
\label{fig:8}
\end{figure}

To proceed further, we may make use of the symmetry (\ref{2.24}) (now
an $U(1)$ gauge symmetry with all $W_i(t)$ being rotations by $\tau_3$).
The analysis appears to be particularly simple in the gauge
where $A_{ij}=0$ across the rungs (Fig.~\ref{fig:8}). 


Then the remaining gauge fields split into the in-phase
and out-of-phase modes
\begin{equation}
A_{\pm}=A_{ij}^{(I)} \pm A_{ij}^{(II)}, \qquad
a_{\pm}=a_i^{(I)} \pm a_i^{(II)}
\end{equation}
(where the superscripts $(I)$ and $(II)$ label the two legs of
the ladder). The dynamics of the gauge fields arises from their
coupling to the spinons and can be found by computing the polarization
diagram (Fig.~\ref{fig:9}), solid lines denote spinons):


\begin{figure}
\centerline{\epsffile{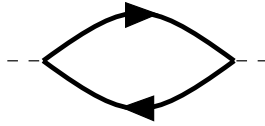}}
\caption{Lowest-order diagram responsible for the dynamics of the
gauge field. Solid lines denote spinon propagators. This diagram gives
mass to the out-of phase mode and 1+1-dimensional QED dynamics for
the in-phase mode.}
\label{fig:9}
\end{figure}

In contrast to the two-dimensional model, where there exist massless
transverse fluctuations of the gauge field, it is not the case
in the ladder. The ladder geometry restricts the transverse
wavevector to two values $k_\perp=0$ and $k_\perp=\pi$
which correspond to the modes $(A_+,a_+)$ and $(A_-,a_-)$
respectively. The out-of-phase modes $A_-$ and $a_-$
(describing the fluctuations of the flux $\varphi$ through the plaquet)
acquire a finite mass of order $J$
and, therefore, can be neglected,
giving only a short-range interactions between holons. 

On the other
hand, the modes $A_+$ and $a_+$ become 1+1 QED gauge fields with
the long-wavelength action
\begin{equation}
S={1\over J^*} \int (\partial_t A_+ + \partial_x a_+)^2 \, dx\, dt
\end{equation}
with $J^*$ of order $J$. This action arises from expanding the
polarization diagram in Fig.~\ref{fig:9}. 
The particular value of $J^*$ depends
on the values of mean-field parameters, and for the typical values
cited above differs from $J$ only by a factor of order unity.

Since the two lower bosonic bands have different isospin, they have
opposite charges with respect to the gauge field $(A_+,a_+)$. In 1+1
dimension, electromagnetic field leads to a confining (linearly
growing with the distance) potential between charges:
\begin{equation}
U(r)=\pm J^* |r|.
\label{2.27}
\end{equation}
In the limit of low hole density,
the bosons will form isospin-neutral dipole pairs (bipolarons)
which produce no field outside each pair. 
Therefore, bipolarons will interact only by short-range forces. 

We may estimate the size of bipolaron by solving a simple quantum-mechanical
problem of two particles interacting
via the potential (\ref{2.27}).
The holon hopping amplitude is of order $t$, therefore the kinetic
energy of bipolaron is of order $t\xi_{pair}^{-2}$. The potential
energy, on the other hand, is of order $J\xi_{pair}$. Thus from
variational principle we find that the size of bipolaron is
$\xi_{pair} \sim (t/J)^{1/3}$ up to a factor of order one. For
our assumption $t/J=3$, this gives $\xi_{pair} \sim 1$, which means
that now the other short-range (repulsive) terms, which we omitted
before, give a comparable contribution. 
Due to the no-double-occupancy constraint, the two species of bosons
must be subject to a substantial on-site repulsion. It increases
the estimated size of the pair by several lattice spacings, 
but does not change the long-range attractive force. Since we
omitted the short-range part of the interaction from the very
beginning, we cannot compute the size of the pair more precisely.
From the numerics \cite{NWS} we know that the characteristic decay length
for the pairing correlation function is about four lattice
spacings, in agreement with our discussion.

\begin{figure}
\centerline{\epsffile{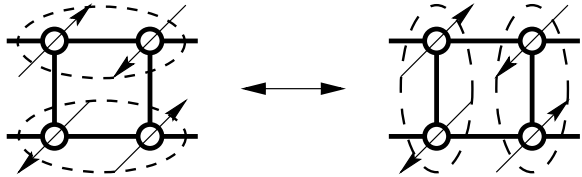}}
\caption{Fluctuating singlets in the spin ladder.}
\label{fig:10}
\end{figure}

A simple classical explanation of the confinement may be obtained from the
picture of fluctuating singlet bonds (somewhat in the spirit of 
\cite{TTR}). We
may think of the spin structure of the Heisenberg antiferromagnet on
the ladder as of spins forming fluctuating singlet bonds with nearest
neighbors (Fig.~\ref{fig:10}).


\begin{figure}
\centerline{\epsffile{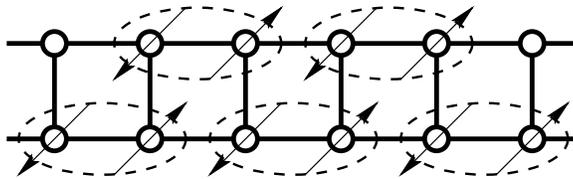}}
\caption{Misplaced singlet bonds between two holes in the t-J ladder.
The energy cost is proportional to the distance between the holes.}
\label{fig:11}
\end{figure}

Once we have a single hole, it leads to the appearance of a
localized spin, which costs a finite energy of order $J$.
When putting two holes, the spins between the holes must form
singlet bonds in a non-favorable way without a freedom
to fluctuate (Fig.~\ref{fig:11}).


Naturally, this string of singlets costs certain energy
$J^* r$, where $r$ is the distance between the holes and
$J^*$ is of order $J$. We believe that this
naive picture gives a correct understanding of the holon
confinement which we derived starting from the mean-field
sF phase.

\section{Luther-Emery liquid and correlation exponents}

In the previous section we have shown that the low-energy excitations
in our model are pairs of holons --- bipolarons --- which are bound by 
long-range confining interaction.
Due to complete screening
of the confining interaction by a single particle in one dimension,
bipolarons interact only via short-range forces, and this recovers
the picture of Luther-Emery liquid of hole pairs proposed earlier
in \cite{TTR}.

In fact, this simple picture is valid at low hole concentration
$\delta\ll\xi_{pair}^{-1}$, when bipolarons do not overlap.
The purpose of this section is to show that as overlap increases,
the charge-density-wave exponents for single-particle and for
pair densities may differ.

For quantitative description of bipolarons we introduce the
three different correlation exponents as follows. 
Let $n(x)=n^{(I)} + n^{(II)}-2\delta$
be the fluctuation of the  number of 
holes on the rung ($x$ is the coordinate along
the ladder, superscripts refer to the two chains). Let further
$n_{pair}=(n^{(I)}-\delta) (n^{(II)}-\delta)$ be 
the fluctuation of the probability that both sites of
the same rung contain holes. Finally, let $\Delta(x)=c^{(I)}_\downarrow
c^{(II)}_\uparrow - c^{(I)}_\uparrow c^{(II)}_\downarrow$ be the
singlet superconducting order parameter on the rung at the position $x$.
Then define the correlation exponents $\alpha_1$, $\alpha_2$ and $\gamma$
by
\begin{equation}
\langle n(x) n(y) \rangle_{2k_F} \propto |x-y|^{-\alpha_1},
\label{3.1}
\end{equation}
\begin{equation}
\langle n_{pair}(x) n_{pair}(y) \rangle_{2k_F} \propto |x-y|^{-\alpha_2},
\label{3.2}
\end{equation}
\begin{equation}
\langle \Delta^\dagger(x) \Delta(y) \rangle \propto |x-y|^{-\gamma},
\label{3.3}
\end{equation}
where $\langle\dots\rangle_{2k_F}$ is the coefficient at $\cos[2k_F(x-y)]$ in
the expansion of the correlation function \cite{H}:
\begin{eqnarray}
\langle n(x) n(y) \rangle =A_0 |x-y|^{-\theta_0} +
A_1 |x-y|^{-\theta_1}\cos[2k_F(x-y)] + \nonumber\\
+ A_2 |x-y|^{-\theta_2} 
\cos[4k_F(x-y)] + \dots
\end{eqnarray}
The correlation function (\ref{3.1}) describes  charge-density-wave (CDW)
ordering, and the correlation function (\ref{3.3}) --- singlet 
superconductivity (SS) ordering.

By $k_F$ we denote the ``Fermi wavevector'' for the holons in the
lower bands, so that $2k_F=2\pi\delta$. For comparison with other works, 
we must remark that $2k_F$ in our notation corresponds to $4k_F$
in the notation of \cite{NWS} and \cite{N}.

In the limit of dilute gas of bipolarons (the doping is much
less than the inverse size of a bipolaron: $\delta \ll \xi_{pair}^{-1}$) we
may describe the low-energy states of the system in terms of bipolaron
creation and annihilation operators (similarly to the
large $|U|$ limit in the attractive Hubbard
model \cite{EL,E}). 
Bipolarons interact repulsively
and, from a naive classical picture, it is likely that the repulsion 
is nearly  hard-core.  In  this limit 
$\alpha_1=\alpha_2$, because on the 
bipolaronic subspace of the total Hilbert space the matrix elements
of the operators $n(x)$ and $n_{pair}(x)$ differ only by a numerical
factor (the probability of the two holons in the pair to occupy the
same rung). For hard-core repulsion $\alpha_{1,2}=2$, since hard-core
bosons can be mapped to free fermions by a Jordan-Wigner-type
transformation, so that density-density correlations coincide with
those of free fermions. From the theory of Luther-Emery liquid it
is known that CDW and SS correlations are described by dual phases,
and the corresponding exponents are therefore reciprocal:
$\gamma=1/\alpha$ \cite{EL,E}. A more detailed discussion of the Luther-Emery
theory of bipolaronic excitations may be found in \cite{TTR}.
 
The opposite limit of highly overlapping bipolarons is more
subtle. In this limit ($\delta\ll (t/J)^{1/3}$) bipolarons can
exchange particles. Exchanging a particle would cost interaction
energy of order $J\delta^{-1}$, while the gain of kinetic energy
would be of order $t\delta^2$. Thus, in this limit we cannot
speak of isolated bipolarons, but rather of two species of
bosons with attraction much smaller than the bandwidth. We
suggest that in this case we may replace the screened long-range
interaction by a short-range one. In the long-wavelength
limit the system may be described as two
Luther-Emery liquids with a weak attraction:
\begin{equation}
H=H_1+H_2+H_{int},
\end{equation}
\begin{equation}
H_i={1\over2\pi}\int dx[v_J (\nabla\varphi_i)^2
+v_N (\nabla\theta_i)^2], \qquad i=1,2,
\end{equation}
\begin{equation}
 H_{int}=-V\int dx \rho_1(x) \rho_2(x),
\label{3.7}
\end{equation}
where $\varphi_i$ and $\theta_i$ are dual phases,
\begin{equation}
[\varphi_i(x), \theta_j(y)]=\delta_{ij} i {\pi\over2}{\rm sign}(x-y),
\label{3.8}
\end{equation}
$v_J$ and $v_N$ are the parameters depending on the short-range properties
of the interaction, and $\rho_i(x)$ are the density fluctuations expressed
by \cite{H}
\begin{equation}
\rho_i(x)={1\over\pi}\nabla\theta_i(x)+2\rho_0 \cos(2k_Fx
+2\theta_i(x)) + {\rm higher~ order~terms}.
\label{3.9}
\end{equation}
The interaction (\ref{3.7}) contains a term proportional to $\cos2(\theta_1(x)
-\theta_2(x))$ and locks the relative phase $\theta_-=\theta_1-\theta_2$.
The only remaining gapless mode is the in-phase fluctuations
($\theta_+=\theta_1+\theta_2$ and $\varphi_+=\varphi_1+\varphi_2$)
corresponding ot propagation of bipolarons.

If we at first approximation neglect higher-order terms in the
density expansion (\ref{3.9}), we find
\begin{equation}
\langle \Delta^\dagger(x) \Delta(y) \rangle \sim
\langle e^{i[\varphi_+(x)-\varphi_+(y)]} \rangle,
\end{equation}
\begin{equation}
\langle n(x) n(y) \rangle_{2k_F} \sim
\langle e^{i[\theta_+(x)-\theta_+(y)]} \rangle,
\end{equation}
\begin{equation}
\langle n_{pair}(x) n_{pair}(y) \rangle_{2k_F} \sim
\langle \nabla\theta_+(x) \nabla\theta_+(y)
e^{i[\theta_+(x)-\theta_+(y)]} \rangle,
\end{equation}
which yields $\alpha_2=\alpha_1+2$, $\gamma=1/\alpha_1$. This
would explain the numerical results of \cite{NWS} who found
$\alpha_2\approx\gamma\approx 2$
(at $\delta=1/8$, $\xi_{pair}\sim 4$).
Our prediction of power-law correlations (\ref{3.1}) also
explains the small, but relatively narrow peak at $2k_F$
in the Fourier transform of $\langle n(x) n(y) \rangle$
in \cite{NWS}.

As it was pointed out by Haldane \cite{H}, in general one must also
include higher-order terms in the density expansion (\ref{3.9}).
These terms proportional to $\cos[m(2k_Fx+2\theta_i(x))]$ with $m>1$
are absent in the free fermion theory (and, consequently, in
hard-core boson theory), but arise as we include interactions
mixing left- and right-moving excitations. They produce terms
proportional to $\exp\, i[m\theta_1(x)-(m\pm1)\theta_2(x)]$ in the
$n_{pair}(x)_{2k_F}$ expansion, and give a contribution to
$\langle n_{pair}(x) n_{pair}(y) \rangle_{2k_F}$ decaying with the
exponent $\alpha_1$ instead of $\alpha_1+2$. This effect
that originally higher-order terms result in leading correlation
exponents is not paradoxical in view of the crossover to the
dilute limit where the binding interaction is strong and the
higher-order corrections to (\ref{3.9}) play a dominating role. 
we suggest that the crossover from $\alpha_2=\alpha_1$ in the
dilute limit to $\alpha_2=\alpha_1+2$ in the weak-coupling
limitis governed by the overlap of bipolarons. Namely,
\begin{equation}
\langle n_{pair}(x) n_{pair}(y) \rangle_{2k_F}=
{A\over(x-y)^{\alpha_1}}+{B\over(x-y)^{\alpha_1+2}}
\label{crossover}
\end{equation}
with relative weights of $A$ and $B$ depending on the average pair
overlap ($A\gg B$ at $\xi_{pair}\delta\ll 1$ and
$A\ll B$ at $\xi_{pair}\delta\gg 1$). The actual
behavior of the coefficients $A$ and $B$ strongly depends
on the short-scale features of the interaction, and
cannot be found in our rude treatment.

The whole discussion of this section is equally applicable to the
negative-U (attractive) Hubbard model. In the low-density
(or large $|U|$) limit the exponents $\alpha_1$ and $\alpha_2$
coincide, while in the small $U$ limit we expect a crossover
(\ref{crossover}) to $\alpha_2=\alpha_1+2$. In other words,
because of screening, the long-range gauge interaction between
holons in one dimension leads to the same behavior 
at large distances as a short-range
attraction.

Finally, we comment on our disagreement with the prediction of
Nagaosa \cite{N} that the correlations 
$\langle n_{pair}(x) n_{pair}(y) \rangle_{2k_F}$ decay
exponentially
(note again that $2k_F$ in our notation 
corresponds to $4k_F$ in the notation of \cite{N}).
The disagreement my be explained from the fact that Nagaosa starts
from two uncoupled chains and treats the interchain
couplings ($t_\perp$ and $J_\perp$) as perturbations. In that picture,
power-law correlations $\langle n_{pair}(x) n_{pair}(y) \rangle_{2k_F}$
will appear as a correction for the nonlinearity of the
spectrum as the coupling increases and approaches the bandwidth.
In contrast to the
weak-coupling approach, our model starts directly from diagonalizing a
strong-coupling Hamiltonian (\ref{2.9}), 
and the correlations (\ref{3.1}) are present from
the very beginning.

\section{Modified d-wave relation on superconducting order parameter}

In this section we verify that our approximation scheme is consistent
with the exact relation for the superconducting order parameter
derived by S.~C.~Zhang for the Hubbard model \cite{Z} (and 
later translated to t-J model in \cite{YZ}). 

Let us define the pairing operator
\begin{equation}
 \Delta_{ij}=c_{i\downarrow} c_{j\uparrow} - 
c_{i\uparrow} c_{j\downarrow}
\end{equation}
and consider the quantity
\begin{equation}
\Delta^{(0)}_{ij}=\langle 2 | \Delta_{ij} | 0 \rangle,
\end{equation}
where $|0\rangle$ and $|2\rangle$ are the ground states with
zero and two holes respectively. Should a superconducting transition
happen, $\Delta^{(0)}_{ij}$ will become the superconducting order
parameter.

From now on, we restrict $i$ and $j$ to be nearest-neighbor sites.
Zhang's results states that on a bipartite lattice in the limit
of zero doping $\Delta^{(0)}_{ij}$ obey the relation
\begin{equation}
\sum_j \Delta^{(0)}_{ij} =0,
\label{4.3}
\end{equation}
where the sum is over the nearest neighbors of the site $i$.
On the two-dimensional square lattice this implies the d-wave
symmetry of pairing; thus we may call Eq.(\ref{4.3}) the
modified d-wave relation.

Below we rederive this result within the $SU(2)$ slave-boson
mean-field approximation. 
For the sake of generality, we
extend our further discussion to the t-J model on an arbitrary
quasi-one-dimensional bi-partite lattice, provided it exhibits 
a spin gap (the most popular examples of this type are even-leg
ladders). Further, assume that the low-temperature mean-field
phase is analogous to the sF phase of the two-leg ladder. Namely,
we require that the $SU(2)$ order parameter may be brought to the
diagonal form (by a suitable choice of gauge):
\begin{equation}
U_{ij}=\pmatrix{\chi_{ij} & 0 \cr 0 & -\chi^*_{ij} }.
\end{equation}
This requirement means that spinons
and holons form doubly degenerate bands 
related by the symmetry of simultaneous time-reversal and
isospin flip. In such a phase a superconductivity may evolve by
pair formation between the holons at the bottom of the two lowest
bands (of course, superconductivity is possible only when stabilized
by inter-ladder interactions, see e.g. \cite{EL}).

For simplicity, let the coupling be isotropic ($t$ and $J$
are the same on all links), as it was assumed in the previous sections.
At the end of this section we shall extend the result to non-isotropic
coupling.

Using our slave-boson representation, we express $\Delta_{ij}$ 
(for nearest-neigbor $i$ and $j$) in
terms of spinons and holons as
\begin{equation}
\Delta_{ij} ={1\over2} \left[
(h^\dagger_i \psi_{2i}) (h^\dagger_j \psi_{1j}) -
(h^\dagger_i \psi_{1i}) (h^\dagger_j \psi_{2j}) \right].
\end{equation}
At low doping, the fermionic part of the correlation function (4.2)
may be replaced by the mean-field order parameters
\begin{equation}
\chi_{ij}=\langle f_{1i}f^\dagger_{1j}+f_{2i}f^\dagger_{2j} \rangle,
\label{4.6}
\end{equation}
and we find
\begin{equation}
\Delta^{(0)}_{ij}={1\over2} \left( \chi_{ij} \langle 2 | b^\dagger_{1i}
b^\dagger_{2j} |0\rangle +
 \chi^*_{ij} \langle 2 | b^\dagger_{2i}
b^\dagger_{1j} |0\rangle \right),
\label{4.7}
\end{equation}
where $|0\rangle$ and $|2\rangle$ now denote the states in the holonic
sector. Let $b_1(k)$ and $b_2(k)$ be the operators destroying
holons at a wave vector $k$ in the two lowest bands (subscripts denote
the isospin). Then the single-pair wave-function $|2\rangle$ has the
form
\begin{equation}
|2\rangle =\int {dk\over 2\pi} \Psi_0(k) b^\dagger_1(k)
b^\dagger_2(-k) |0\rangle,
\end{equation}
where $\Psi_0(k)$ is the relative wave function of the two holons
in a pair. Eq.(\ref{4.7}) becomes
\begin{equation}
\Delta^{(0)}_{ij}=\int {dk\over 2\pi} \Psi_0(k) \Delta_{ij}(k),
\end{equation}
where
\begin{eqnarray}
 \Delta_{ij}(k) ={1\over2}\left( \chi_{ij} \langle 0|b_2(-k)b_1(k)
b^\dagger_{1i} b^\dagger_{2j} |0\rangle +
\chi^*_{ij} \langle 0|b_2(-k)b_1(k)
b^\dagger_{2i} b^\dagger_{1j} |0\rangle \right)= \nonumber\\
={1\over2}\left( \chi_{ij} \langle 0|b_2(-k)b^\dagger_{2j} |0\rangle
\langle 0|b_1(k)b^\dagger_{1i} |0\rangle +
 \chi^*_{ij} \langle 0|b_2(-k)b^\dagger_{2i} |0\rangle
\langle 0|b_1(k)b^\dagger_{1j} |0\rangle \right). 
\end{eqnarray}
Since $b_1(k)$ and $b_2(k)$ are related by the time-reversal symmetry
(accompanied by a gauge transformation),
$\langle 0|b_2(-k)b^\dagger_{2i} |0\rangle = (-1)^i
\langle 0|b_{1i} b^\dagger_1(k) |0\rangle$, and for nearest-neghbor
sites $i$ and $j$
\begin{eqnarray}
\Delta_{ij}(k)={1\over2}\left( (-1)^j
 \chi_{ij} \langle 0|b_{1j} b^\dagger_1(k) |0\rangle
\langle 0|b_1(k)b^\dagger_{1i} |0\rangle + (-1)^i
 \chi^*_{ij} \langle 0|b_{1i}b^\dagger_1(k) |0\rangle
\langle 0|b_1(k)b^\dagger_{1j} |0\rangle \right) \nonumber\\
=(-1)^i \langle k | {1\over2}(-\chi_{ij} b^\dagger_{1i}b_{1j}
+\chi^*_{ij} b^\dagger_{1j}b_{1i}) | k \rangle, 
\end{eqnarray}
where $|k\rangle$ is a single-holon plane wave created by $b^\dagger_1(k)$.
The state $|k\rangle$ is an eigenvector of the free Hamiltonian
proportional to the bosonic part of Eq.(\ref{2.9})
\begin{equation}
H_0={1\over2}\sum_{\{ij\}}(b^\dagger_{1i} \chi_{ij} b_{1j}
+ b^\dagger_{1j} \chi^*_{ij} b_{1i})
\label{4.12}
\end{equation}
(with the sum performed over nearest-neighbor pairs of sites).
Therefore
\begin{equation}
\sum_j \Delta_{ij}(k)=(-1)^i \langle k| 
\big[ b^\dagger_{1i} b_{1i},H_0\big]
|k \rangle =0,
\end{equation}
where the sum is over the nearest neighbors of the site $i$.
This immediately implies the result (\ref{4.3}).

The above derivation holds also at a non-zero temperature (with
ground-state expectation values replaced by thermal averages). However,
it is strictly limited to zero doping: at finite doping Eq.(\ref{4.6}) is
no longer valid.
A remarkable feature of the relation (\ref{4.3}) is its independence
of the coupling parameters $t$ and $J$.

For the two-leg ladder the modified d-wave relation (\ref{4.3}) turns into
\begin{equation}
\Delta_\perp=-2\Delta_\parallel.
\end{equation}
This result agrees with the available numerical results 
\cite{NWS,SRZ}.
It would provide a good test for possible numerical models
on ladders with a higher number of legs.

The d-wave relation (\ref{4.3}) may be easily extended to a non-isotropic
coupling. In fact, the whole slave boson mean-field theory may be 
rederived for arbitrary coupling constants $t_{ij}$ and $J_{ij}$,
differing at different links. One just needs to replace $t$ and $J$
in Eqs.~(\ref{2.6}) -- (\ref{2.9}) by $t_{ij}$ and $J_{ij}$. 
The whole argument
of this section may be repeated downto Eq.(\ref{4.12}) which we must
now replace by the Hamiltonian 
\begin{equation}
H_0={1\over2}\sum_{\{ij\}}t_{ij}(b^\dagger_{1i} \chi_{ij} b_{1j}
+ b^\dagger_{1j} \chi^*_{ij} b_{1i}).
\end{equation}
Finally, this leads to the following generalization of the d-wave
relation (\ref{4.3}):
\begin{equation}
\sum_j t_{ij} \Delta^{(0)}_{ij} =0.
\label{4.16}
\end{equation}
This equation implies that as the hopping on a link increases,
the weight of the superconducting oreder parameter on this link
decreases. Of course, Eq.(\ref{4.16}) may also be derived exactly for the
Hubbard model (with the site- and link-dependent $U$ and $t$) by the
method of \cite{Z}.

\section{Conclusion}

We presented the low-energy effective theory for charge
excitations in two-leg t-J ladder, based on the mean-field
treatment of spin degrees of freedom. We found that the
$SU(2)$ slave-boson formalism predicts bipolaronic picture
of charge excitations, as expected from earlier analytic
and numerical works. While capturing well the low-energy
physics, our approximation is not reliable for spin and
single-electron excitations which have a gap of order $J$.
In the framework of the model developed in the paper,
single-hole excitations may be constructed as holon-spinon
pairs bound by confining gauge-field interaction. 
The similarity of the boson-fermion spectrum
in Fig.~\ref{fig:7} to the single-hole spectrum found numerically
in \cite{TTR} makes this possibility very appealing.
However, spin structure of the ladder was included
in the lowest mean-field order, and to restore correctly
spin excitations requires a more elaborate treatment.

\acknowledgements

We wish to thank Christopher Mudry for illuminating discussions
of one-dimensional systems and for many helpful remarks. We are
grateful to N.~Nagaosa, X.-G.~Wen, and D.~H.~Kim for useful
discussions. This work is supported by NSF under the MRSEC program,
DMR 94-0034.

\end{document}